\documentclass[%
reprint,
superscriptaddress,
 amsmath,amssymb,nofootinbib,
 aps,
 twocols
]{revtex4-1}

\usepackage[title]{appendix}
\usepackage{verbatim}
\usepackage{graphicx}
\usepackage{dcolumn}
\usepackage{bm}
\usepackage{mathtools}
\usepackage{float}
\makeatletter
\let\newfloat\newfloat@ltx
\makeatother
\usepackage{algorithm} 
\usepackage{algpseudocode}  

\usepackage{float}                                    
\usepackage{graphicx} 
\usepackage[font=small,labelfont=bf,justification=raggedright,format=plain]{caption}
\usepackage{hyperref}                                 
\usepackage[usenames, dvipsnames]{color}
\usepackage{color}
\usepackage{listings}
\usepackage{enumerate}

\definecolor{dkgreen}{rgb}{0,0.6,0}
\definecolor{gray}{rgb}{0.5,0.5,0.5}
\definecolor{mauve}{rgb}{0.58,0,0.82}

\begin{document}


\title{Information-theoretical measures identify accurate low-resolution representations of protein configurational space}


\author{Margherita Mele}
\affiliation{Physics Department, University of Trento, via Sommarive, 14 I-38123 Trento, Italy}
\author{Roberto Covino}%
\affiliation{Frankfurt Institute for Advanced Studies, 60438 Frankfurt am Main, Germany}
\author{Raffaello Potestio}%
 \email{raffaello.potestio@unitn.it}
\affiliation{Physics Department, University of Trento, via Sommarive, 14 I-38123 Trento, Italy}
\affiliation{INFN-TIFPA, Trento Institute for Fundamental Physics and Applications, I-38123 Trento, Italy}

\date{\today}

\begin{abstract}
A steadily growing computational power is employed to perform molecular dynamics simulations of biological macromolecules, which represents at the same time an immense opportunity and a formidable challenge. In fact, large amounts of data are produced, from which useful, synthetic, and intelligible information has to be extracted to make the crucial step from knowing to understanding. Here we tackled the problem of coarsening the conformational space sampled by proteins in the course of molecular dynamics simulations. We applied different schemes to cluster the frames of a dataset of protein simulations; we then employed an information-theoretical framework, based on the notion of {\it resolution and relevance}, to gauge how well the  various clustering methods accomplish this simplification of the configurational space. Our approach allowed us to identify the level of resolution that optimally balances simplicity and informativeness; furthermore, we found that the most physically accurate clustering procedures are those that induce an ultrametric structure of the low-resolution space, consistently with the hypothesis that the protein conformational landscape has a self-similar organisation. The proposed strategy is general and its applicability extends beyond that of computational biophysics, making it a valuable tool to extract useful information from large datasets.
\end{abstract}

\maketitle

\begin{figure}[t]
\centering
\includegraphics[width=\columnwidth]{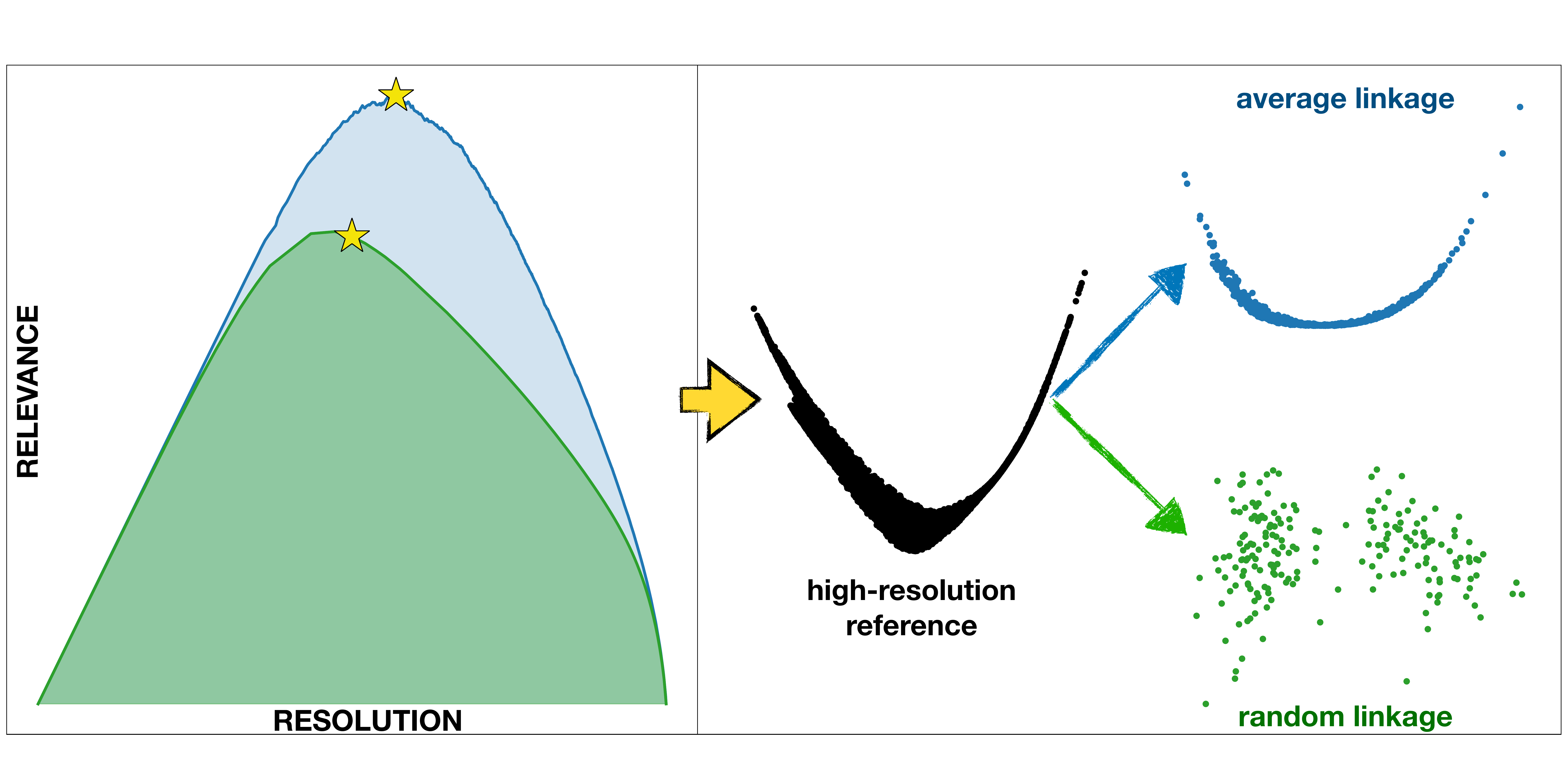}
\caption*{\footnotesize Table-of-contents entry. Higher values of relevance are associated to clustering methods producing structurally accurate coarse representations of a protein's conformational space.}
\label{fig:toc}
\end{figure}

\section{Introduction}

A celebrated quote attributed to Aristotle states that ``the whole is more than the sum of its parts''. This statement effectively encapsulates the defining characteristic of complex systems, whose global properties generally cannot be traced back to those of their individual constituents, but rather {\it emerge} from the interplay of the latter.

Among those systems that most clearly show this behaviour, a prominent example is represented by biological macromolecules such as proteins: these, being composed of several hundreds of interacting atoms, display a rich and sophisticated phenomenology unfolding over broad ranges of length and time scales, which cannot be naively predicted or anticipated from the knowledge of their structure. In order to generate, inspect, and comprehend the properties and behaviour of these systems, computational, {\it in silico} methods have been developed, most notably molecular dynamics~\cite{karplus1990molecular,gonzalez2011force,pan2016demonstrating,adcock2006molecular} (MD) simulations, that serve the purpose, among others, of sampling the conformational space of the molecule. Once a dataset of sampled conformations, or frames, is available, however, one faces the problem of extracting useful and intelligible information out of it, separating the relevant feature from the irrelevant detail.

This task can be carried out through dimensionality reduction~\cite{tribello2019using} or clustering schemes. These methods rely on some notion of similarity - usually a structural similarity - between distinct conformations to group together those whose differences are negligible, while a much larger discrepancy exists from other frames or groups of frames. It might appear desirable to devise these clustering schemes taking advantage of a preexisting knowledge about the system, in order to steer the algorithm towards physically sensible partitions of the sampled conformational space. It can be the case, however, that an undesired bias is introduced in the process, with potentially detrimental consequences for the interpretation of the results; alternatively, one might hope for a completely undirected procedure~\cite{glielmo2021unsupervised,noe2017collective,glielmo2021ranking}, so as to let the system itself dictate how to cluster its data points, and allow the intrinsic organisation of the conformational space to emerge.

A recently developed information-theoretical approach, the resolution-relevance framework~\cite{battistin2017learning}, holds the promise to carry out this task of identifying intrinsically informative low-dimensional representations of the system in an unbiased manner. This approach relies on distinct measures of the information content of a dataset to group the instances of the latter in a way that optimally separates information from noise, and allows the extraction of the largest amount of information about the generative process that underlies the data points. The method, however, operates on the basis of a predefined classification procedure, whose impact cannot be neglected in the assessment of the resulting partition's quality and physical soundness: in fact, the values of these information metrics for a given arrangement of the data points in clusters only make sense relative to the strategy employed to perform the grouping.

In this work, we tackle the issue of investigating if, and to what extent, different strategies to carry out the clustering of protein MD trajectory frames affect the intrinsic quality of the resulting partitions, and if the resolution-relevance framework can be employed to make sense of these results. We apply this strategy to a dataset of $12$ structurally dissimilar proteins as well as to a specific case study, making use of $7$ different linkage criteria for the clustering. Our results support the hypothesis that the resolution-relevance analysis can select those linkage methods giving rise to low-resolution representations of the protein conformational space that reproduce the high-resolution reference with the highest degree of fidelity; furthermore, we propose that this capacity of performing a sensible clustering is a direct consequence of the clustering method being capable of preserving the intrinsically hierarchic structure and ultrametricity of the protein conformational space.

\section{The relevance-resolution framework and the impact of the underlying clustering method}
\label{sec:RR_framework}

The resolution-relevance framework, or critical variable selection, is a recently developed method~\cite{grigolon2016identifying} for identifying important variables without any prior knowledge of, or assumption on, their nature. The idea at the heart of the approach is that the information on the generative model that underlies the elements of an empirical sample is contained in the distribution of their frequencies, that is to say, in the number of times different outcomes occur in the data set. It can be shown~\cite{marsili2013sampling, haimovici2015criticality} that the entropy of the outcome distribution, dubbed {\it resolution}, quantifies the overall information content of the sample, while the entropy of the frequency distribution, dubbed {\it relevance}, measures the amount of important information. In this section we provide a synthetic review of this approach, specialising the formulation for the application in the context of computational biophysics.

The output of a molecular dynamics simulation consists of a collection of configurations, or frames, $\hat{s}=\big(s^{(1)},...,s^{(M)}\big)$; these can be thought of as the realisations of a stochastic sampling process, where each element takes the values of one of the possible system states $s=\big( s_1,....s_n\big)$, with $n\gg M$. In spite of absolute structural differences, two distinct configurations might result equivalent for a practical purpose; for example, if the relative position of a few atoms in two frames differs by less than a given tolerance, they might be considered essentially equivalently representative of the same overall organisation of the molecule. In analysing the outcomes of a simulation it is thus crucial to filter out redundant details by grouping together structures that can be safely associated to the same state; hence, one has to perform a {\it clustering}.

The most trivial level of clustering consists in identifying each frame as a distinct cluster (assuming that no pair of exactly identical configurations exists in the sample). Such a representation clearly allows the highest level of detail in the description of the dataset, but it bears no use in making sense of it; the number of clusters thus has to be reduced, and frames that in principle describe distinct structural organisations have to be grouped together if their distance (as quantified by an appropriate measure) takes values below a predefined threshold. In so doing, the number of clusters is reduced from $K = M$ to values $K < M$, which correspond to increasingly less resolved representations of the system's configuration space.

For each partition of the dataset it is possible to compute the corresponding values of the aforementioned resolution and relevance. Resolution is defined as (note that we employ logarithms in units of $M$, or {\it Mats}, so that $\log_M M = 1$):
\begin{equation}
    H[{s}]=-\sum_{s=1}^{K} \frac{k_s}{M} \log_M\frac{k_s}{M}
    \label{eq:rel}
\end{equation}
where $k_s$ is the number of frames associated to the cluster with label $s$, and $k_s/M$ is the empirical probability that a randomly chosen frame from the data set belongs to cluster $s$. The normalisation condition $\sum_s k_s = M$ ensures that $k_s/M$ is indeed at most unity.

Since all frames in a cluster are indistinguishable at the level of detail employed, the lowest resolution value $H[{s}] = 0$ is obtained when all frames are gathered in the same cluster; similarly, the largest value $H[{s}] = \log_M M = 1$ is attained when each frame is a singleton cluster. Both extremes are equally little informative: on one hand, when the resolution is too low, potentially different conformations are grouped in the same cluster; on the other hand, discriminating all $M$ states as distinct is equivalent to associate to each of them the same probability, which does not provide useful information to infer the underlying generative process. Hence, resolution alone is not sufficient to pinpoint an optimal level of detail at which the system should be inspected, and a second measure has to be employed to this end. Such measure is the relevance $H[{k}]$, given by:
\begin{equation}
    H[{k}]=-\sum_{k} \frac{km_k}{M} \log_M\frac{km_k}{M}
    \label{eq:res}
\end{equation}
where $m_k$ is the number of outcomes $s$ for which $k_s = k$, and $km_k/M$ is the empirical probability that a frame chosen at random from the data set is associated to a state with (un-normalised) frequency $k$.

The relevance is null for both extreme values of the resolution: in the case $H[{s}] = 0$ all frames are in the cluster with $k = M$, which gives $m_M = 1,\ m_{k \neq M} = 0$, and hence $\frac{km_k}{M} \ln_M \frac{km_k}{M} = 0\ \forall\ k$; in the case $H[{s}] = 1$ all clusters only contain one frame, hence $m_1 = M,\ m_{k \neq 1} = 0$ and $\frac{km_k}{M} \ln_M \frac{km_k}{M} = 0\ \forall\ k$ as well.

As the relevance is nonnegative and equal to zero at the extremes of the resolution range, it follows that there must be one representation, with intermediate resolution and positive relevance value, that more than the others allows an informative characterisation of the underlying probability distribution~\cite{song2018resolution}; hence, the relevance as a function of the resolution has to have a maximum. The partitions at the right of this maximum are in what is called the {\it under-sampling regime}, $M \ll n$, in which the statistics of the data is relatively poor and several frames associated to distinct states can happen to appear the same number of times. For a given value of the resolution in this region, those partitions that maximise the relevance - the {\it most informative samples} - feature a frequency distribution that follows a power law, $m_k\sim k^{-\mu -1}$ with $\mu > 0$, such that each value of the frequency is associated to a distinct number of clusters. In particular, the partition for which the quantity $H[{s}] + H[{k}]$ is the largest has $\mu = 1$: this corresponds to Zipf's law, $m_k\sim k^{-2}$, which is associated to the point of optimal tradeoff between parsimony of the representation (low resolution) and its informativeness (high relevance)~\cite{cubero2018minimum, cubero2019statistical}. This is the case, for example, for the frequency of the words in a language~\cite{zipf2013selected}, and the spike patterns of neuron populations~\cite{tkavcik2015thermodynamics}.

\begin{figure}[t]
\centering
\includegraphics[width=0.49\textwidth]{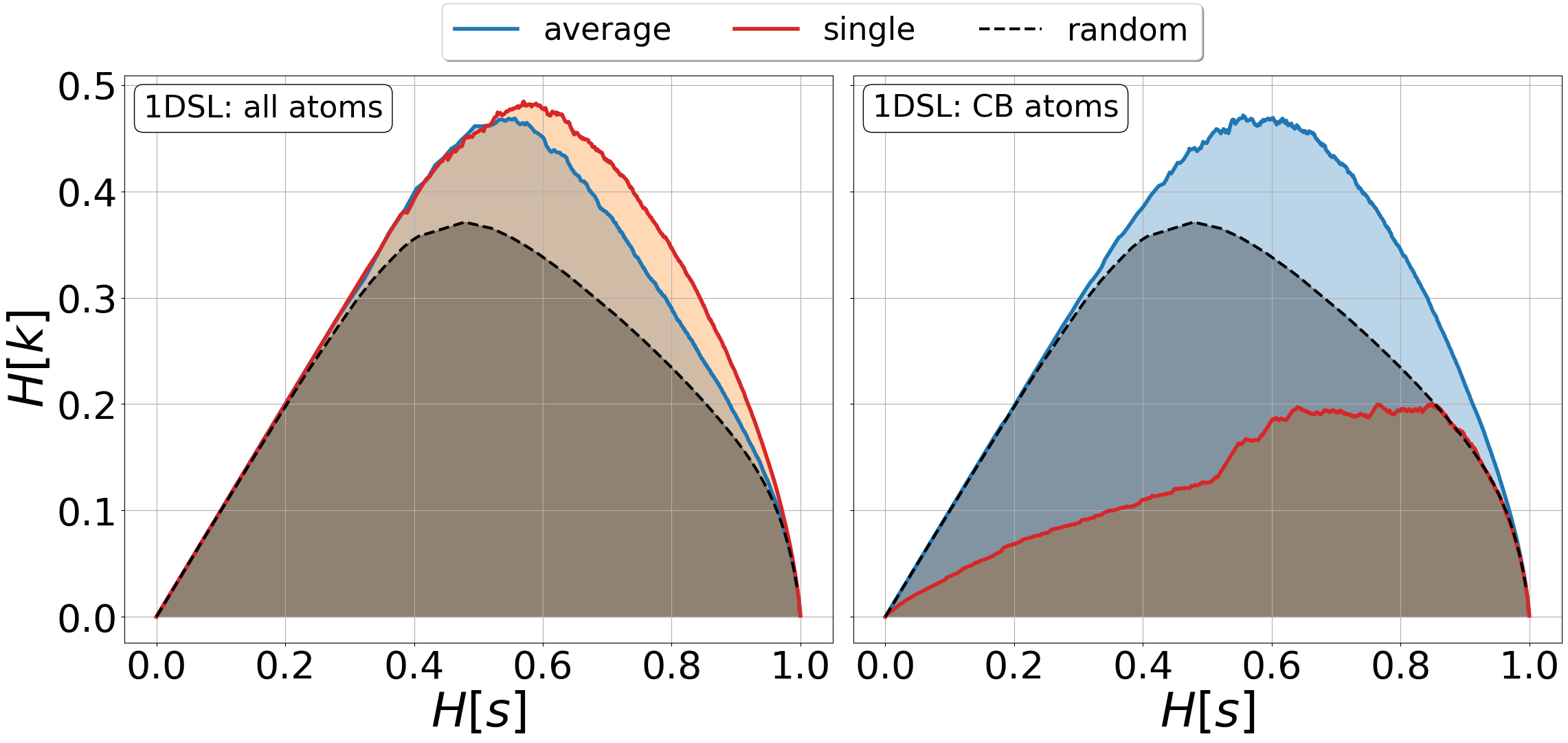}
\caption{\footnotesize Relevance-resolution curves obtained partitioning the simulation data of protein 1DSL with two clustering protocols, average linkage and single linkage. The two panels differ by the atom selection adopted: \textit{all atoms} on the left and \textit{C$_\beta$ atoms} on the right. In both panels the \textit{random} curve is also present, obtained by randomly partitioning the structures into groups (see \hyperref[subsec:random]{Methods}). Each $(H[s],H[k])$ point corresponds to a fixed number of clusters in which the frames of the whole trajectory are grouped. All curves show the expected characteristic trend: zero relevance at the lowest (all frames in a cluster) and highest (every frame in a single cluster) resolution value.}
\label{fig:MSR}
\end{figure}
In a context of complete ignorance, i.e. in absence of any information about the data except their empirical probability $k/M$ based on some pre-defined classification, the frequency is the only label that can be employed to distinguish between frames in distinct states~\cite{cubero2018minimum, cubero2019statistical}. The frequency thus constitutes a {\it minimally sufficient representation}, which, in absence of additional information about the data, allows one to write the resolution $H[s]$ as:
\begin{equation}
    H[s] = H[k] + H[s|k]
    \label{eq:resreldef}
\end{equation}
where the information content of a given partition in states $s$ is decomposed in the relevance and a noise term, $H[s | k]$. The latter is the larger for a partition based on the frequency than for any other partition, and it constitutes a measure of the degeneracy of the distinct classifications that produce the same frequency distribution~\cite{cubero2019statistical}. In fact, different classifications that preserve the number of elements in each state are fully equivalent, as relevance and resolution solely depend on the partition's combinatorics: that is, \textit{how many clusters} are there and \textit{how many elements} there are in each cluster; both quantities are blind to \textit{which} elements are included in a given cluster.

This is a crucial aspect, which shows that the implication \textit{high relevance therefore informative representation} is not necessarily true. Indeed, even a random clustering of the system may produce partitions with high relevance values but devoid of any significant information (\hyperref[fig:MSR]{Fig.\ref{fig:MSR}}). Consequently, the relevance-resolution framework cannot be used on its own, e.g. looking for the partition that, at a given value of the resolution, maximises the relevance: rather, it has to be associated with a sensible strategy to group elements in clusters, based on the physical (geometric, structural, energetic...) properties of the sample. The most informative representation is thus the one maximising the relevance {\it compatibly with the constraints imposed by the clustering algorithm}.

In the case of molecular structures, the clustering procedure relies on the specific functions defining the inter-frame and inter-cluster distance (\hyperref[fig:MSR]{Fig.\ref{fig:MSR}}). The former defines the property in terms of which the similarity of two configurations is quantified (structure, compactness, energy, etc.), while the latter is the metric employed to measure the distance between clusters. The latter, which is referred to as {\it linkage criterion}, thus determines the protocol employed for clustering, and different choices result in different partitions of the system. The ability of a given protocol to return a meaningful partitioning naturally depends on the specific dataset under examination. For example, in single linkage the similarity of two clusters is equivalent to that of their most similar members; this protocol is effective in identifying compact and separate clusters, but it is strongly subject to the \textit{chaining effect}: two close-by points can form a bridge between two clusters, causing them to merge and resulting in an elongated cluster. Without any prior characterisation of the explored configurational space, the goodness of the partition can be assessed only {\it a posteriori}.

In this work, we employed various linkage criteria and investigated the most informative partitions obtained with each of them, with two objectives: first, to identify those linkage methods that are most appropriate for a meaningful and physically sound clustering of molecular structure data; second, to infer general properties of the configurational space explored by proteins in molecular dynamics simulations. In the next section we report the results of our study; the list and definitions of the employed distance measures and linkage criteria are provided in \hyperref[app:linkage]{Appendix A}.

\begin{figure*}[t]
  \includegraphics[width=\textwidth]{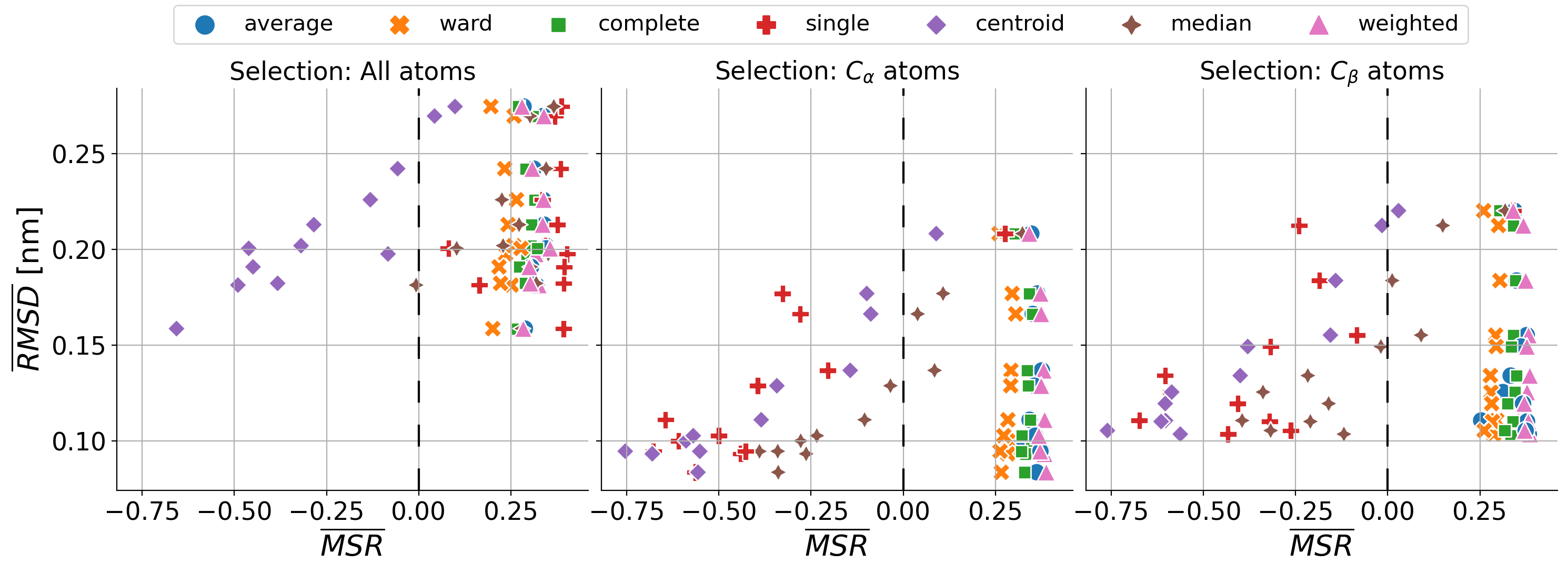}
  \caption{\footnotesize Performance of clustering algorithms, as quantified by the relative $\overline{MSR}$, related to the protein mobility, given by the mean value of the RMSD matrix ($\overline{RMSD}$). Each point corresponds to one of the $12$ proteins of the dataset. The three panels differ by the atom selection adopted for clustering: \textit{all atoms} on the left, \textit{C$_\alpha$ atoms} in the middle, and \textit{C$_\beta$ atoms} on the right. The colours and the markers identify the linkage criterion used in the clustering procedure as shown in legend. }
  \label{fig:RMSD-MSR}
\end{figure*}

\section{Results}

We investigated the impact of choosing seven different linkage methods on the relevance-resolution curves obtained clustering protein structures sampled in molecular dynamics simulations (see \hyperref[sec:Methods]{Methods}). To this end, we employed two set of systems: a dataset of $12$ structurally distinct molecules, and a specific protein, adenylate kinase~\cite{ionescu2019adenylate,ake-struct}, which was taken as a case study. The similarity of configurations was quantified in terms of the root mean square distance (RMSD) between frame pairs, and we analysed configurations at three levels of structural resolution: all-atom, which accounts for all heavy atoms of the molecule; alpha carbon (C$_\alpha$); and beta carbon (C$_\beta$). The last two consist in reduced representations with a single coarse-grained site per amino acid. Resolution-relevance curves were computed in all aforementioned cases, however the overall informativeness of a given representation was assessed in terms of the multi-scale relevance~\cite{cubero2020multiscale} (MSR), i.e. the area under the relevance curve: this is a global measure of how a specific clustering scheme performs at all levels of resolution.

The data in \hyperref[fig:MSR]{Fig.\ref{fig:MSR}} suggest that it is possible to characterise the ability of a clustering method to identify informative partitions across various levels of structural detail. Indeed, it can be seen that, although all the curves in the figure were obtained by clustering the same trajectory frames, the MSR values obtained are rather different. In particular, the selection of atoms drastically influences the MSR value obtained through a given clustering protocol. The results of this analysis, carried out for $7$ linkage criteria and $12$ proteins at varying levels of structural resolution (all, C$_\alpha$, and C$_\beta$ atoms), are summarised in \hyperref[fig:RMSD-MSR]{Fig.\ref{fig:RMSD-MSR}}, where the deviation of the MSR relative to the random reference ($\overline{MSR}$, see \hyperref[eq:MSR-norm]{Eq.\ref{eq:MSR-norm}}) is plotted against the mean value of the RMSD matrices used for the clustering procedure. For some linkage criteria, the $\overline{MSR}$ values are always positive (i.e. larger than the random value $MSR_R$) regardless of the system and structural selection employed; for other methods, the performance depends on the system or its representation. In particular, the plot shows that the performance of some methods correlates with the mobility of the protein as quantified by the average RMSD. The linkage criteria can be divided in two groups: those for which the MSR value correlates with the mean value of the RMSD matrix (`centroid' and `median') and those for which it does not (`average', `complete', `weighted' and `ward'). Consequently, it can be argued that the former are less effective and reliable than the latter: in fact, their ability to identify simplified but meaningful representations strongly depends on the specific data under examination, and their applicability is restricted to high-resolution (all-atom) descriptions. This represents a substantial shortcoming, in that these methods are not adequate for the analysis of all-atom data in less detailed terms.

The single linkage criterion shows a peculiar pattern, and deserves to be discussed separately. Here, the distance between clusters is the minimum pairwise distance between their elements. Hence, in the all-atom case, as the matrix elements are widely spread, this algorithm manages to form differently-populated clusters. Conversely, the C$_\alpha$ and C$_\beta$ selections implement a coarse-graining that ``blurs'' the structural differences from the outset; therefore, the algorithm tends to form highly populated clusters by putting together even frames that are relatively different from each other (chaining effect), and provides a rather uninformative representation of the system. 

It can be shown~\cite{milligan1979ultrametric} that some of the hierarchical clustering algorithms induce a monotonic hierarchy, i.e. the values in the inter-cluster distance matrix increase monotonically during agglomerative clustering. Algorithms that induce a monotonic hierarchy lead to an ultrametric in the cluster space~\cite{johnson1967hierarchical}: this implies that the metric distance satisfies an inequality stronger than the triangular one~\cite{jardine1971mathematical,fushing2013multi}. In our analysis, it turns out that clustering protocols that satisfy these qualities coincide with those showing a consistently positive $\overline{MSR}$; the only exception to this trend is single linkage, which, although inducing an ultrametric in cluster space, still shows negative $\overline{MSR}$ values when coarse-grained representations of the system are employed. In this case, however, the clustering protocol is severely limited by the chaining effect, plausibly producing uninformative partitions of the system and consequently obtaining a lower or comparable MSR value with respect to the random case. Taken together, these results suggest that protein structures sampled in the course of a molecular dynamics simulation populate the configurational space according to an ultrametric structure, which is consistent with the self-similar organisation of the free energy landscape observed in previous works~\cite{pontiggia2007anharmonicity,volkhardt2022estimating,pandya2018energy,henzler2007hierarchy,henzler2007dynamic}; additionally, the MSR appears to be capable of capturing, in a parameter-free and unbiased manner, the effectiveness of a clustering method in finding informative representations of a biomolecule's configurational variability at different scales of resolution, in that MSR correlates with the method's capacity to preserve the ultrametric structure of the reference configurational space.

Since relevance and resolution are not sensitive to the features of the elements gathered in the clusters and their relative similarity, it is crucial to validate {\it a posteriori} that partitions with a higher relevance are indeed more informative than the others. This task can be achieved through dimensionality reduction techniques~\cite{DiffMap1,lafon2006diffusion,nadler2006diffusion}, which project the high-dimensional trajectory of the molecule in Cartesian coordinate space onto a low-dimensional manifold of collective coordinates. We thus performed a comparison of the distribution of points (frames or cluster centroids) in the space spanned by the first two, most collective diffusion coordinates (DC) obtained from the \textit{high-resolution} (HR) or \textit{low-resolution} (LR) representation of the system (see \hyperref[sec:Methods]{Methods}). It is reasonable to expect that a meaningful partition gathers, in the same cluster, frames close in HR diffusion space, and that the distribution of centroids resembles the HR distribution, thus allowing the same information to be extracted. In order to assess and compare the goodness of partitions we resort to the decomposition of the covariance matrix in its inter- and intra-state contributions (see Eqs. \ref{eq:cov1}-\ref{eq:cov3} in the \hyperref[sec:Methods]{Methods} section). In fact, a key property of an informative LR representation of a system is to capture more information in the retained data than what is left in the discarded ones; we thus expect that the trace of the inter-cluster covariance will be significantly higher than the intra-cluster one.

\begin{figure*}[ht]
  \includegraphics[width=\textwidth]{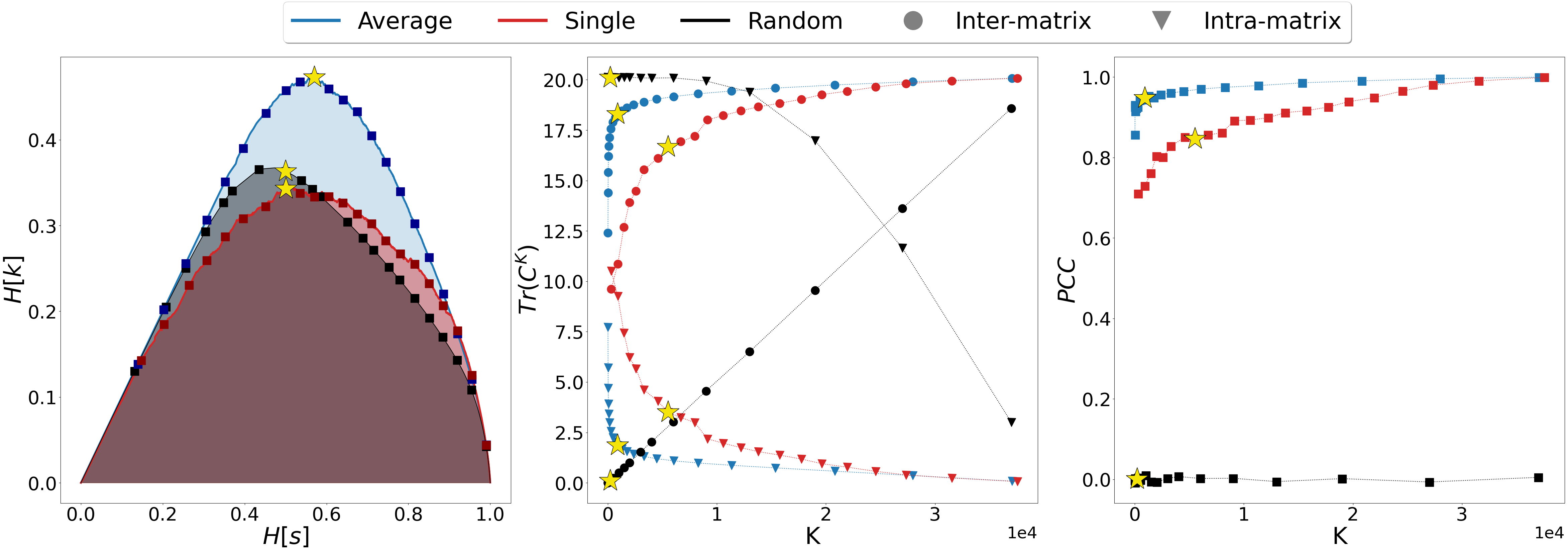}
  \caption{\footnotesize Comparison of the different LR representations obtained from clustering the MD trajectory of adenylate kinase using the average linkage (blue), single linkage (red), and random clustering (black). The system was analysed using a coarse representation in which only C$_\beta$ atoms are considered. Left pane: relevance-resolution curves drawn by the different protocols; the square points mark the $22$ low-resolution representations under examination. Middle pane: traces of the inter-state (circles) and intra-state (triangles) correlation matrix plotted against the number of clusters $K$. Right pane: Pearson correlation coefficient (PCC) between the first DCs in the HR and LR representation, plotted against the number of clusters $K$. The yellow star in each of the graphs indicates the representation that maximises the relevance for the corresponding clustering method.}
  \label{fig:1ake-PCC-cov}
\end{figure*}

We thus proceeded to investigate in greater detail the relationship between linkage method and informativeness of the resulting low-resolution representation of a protein's conformational space. To this end, we focused on a specific case study, that of adenylate kinase: the configurations obtained from a $800 \;ns$ long simulation, reduced to the positions of the sole C$_\beta$ atoms, were grouped with the single, average, and random clustering methods at various levels of cluster numbers, corresponding to $22$ different, resolution values in the range (0,1); the spacing in resolution between the first $7$ representations is $\sim0.06$, for the remaining ones is $\sim0.04$. For each LR partition we computed, and reported in \hyperref[fig:1ake-PCC-cov]{Fig.\ref{fig:1ake-PCC-cov}}, the resolution-relevance plots (left pane), the trace of the inter-state and intra-state matrix (middle pane), and the value of the Pearson correlation coefficient (PCC) between the first diffusion coordinates in HR and LR space (right pane); the last two sets of quantities are plotted against the number of clusters $K$ employed in the representation.

\begin{figure*}
    \centering
    \includegraphics[width=\textwidth]{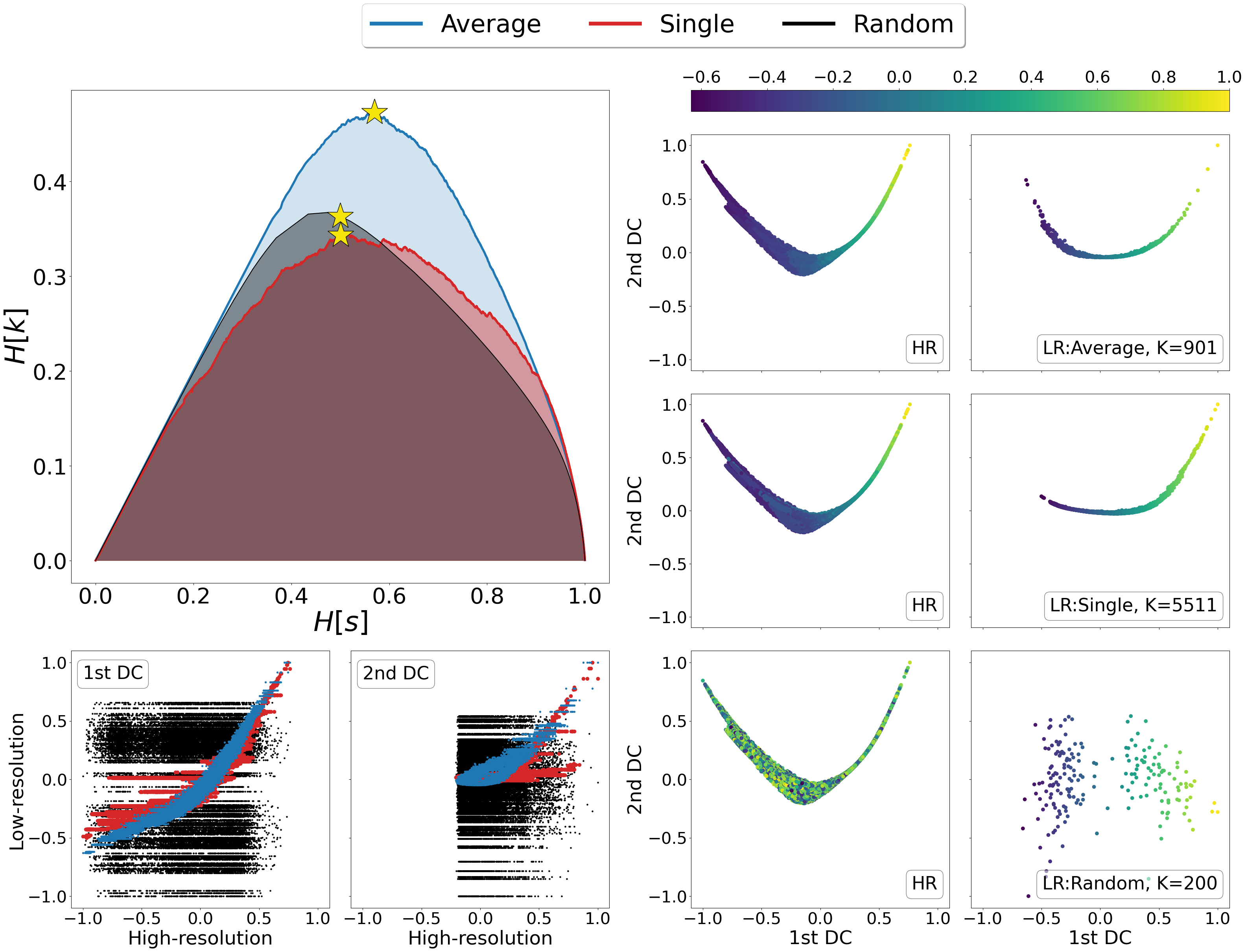}
    \caption{\footnotesize Detailed analysis of the highest-relevance partitions of adenylate kinase trajectory data obtained with single linkage, average linkage, and random clustering. Upper left-hand panel: relevance-resolution curves obtained from clustering the MD trajectory of adenylate kinase {\it via} average linkage (blue), single linkage (red), and random clustering (black). The system was analysed using a coarse representation in which only the C$_\beta$ atoms are considered. The yellow stars indicate the position, on the corresponding curve, of the low-resolution representations maximising the relevance obtained by partitioning the system into a $K$ clusters. Right panel: trajectory frames or cluster points projected onto the space spanned by the first two diffusion coordinates. The panel's left column shows the diffusion space resulting from high-resolution (HR) representation where each point is a frame of the MD simulation, and the distance between points is given by the RMSD computed on the C$_\beta$ atoms; the right column shows the two-dimensional diffusion space resulting from the low-resolution (LR) representation, where each point is the centroid of a cluster and the distance between points is the linkage measure that produced the partition for the information-driven clustering (average linkage in the first row and single linkage in the second row), or the RMSD between the clusters centroid for the random clustering (third row). In both high- and low-resolution space, the points are coloured according to the value taken by the first diffusion coordinate in the low-resolution space. Lower left-hand panel: scatter plot of the first (left) and second (right) diffusion coordinates of the HR space plotted against the corresponding coordinates of the LR space; in each graph, we report the points of the representations obtained through random clustering (black), single linkage (red), and average linkage (blue). Note that the compared LR representations display close resolution values ($H[s]$) but significantly different numbers of clusters ($K$).}
    \label{fig:1akeDC}
\end{figure*}

It is possible to observe that, at a fixed level of resolution, the LR representations obtained through the average linkage are simpler and more informative, as the corresponding number of clusters $K$ is lower and the relevance value higher than those obtained by single linkage; the latter also produces a relevance curve that lies very close to that of the random partition. This hierarchy in performance is also confirmed by the trends of the trace of the covariance matrices and PCC: already with a small number of clusters ($K \sim 10$) average linkage identifies LR representations in which the inter-cluster contribution is significantly higher than the intra-cluster one. In contrast, at the same number of clusters the single linkage algorithm produces partitions for which the two terms are of the same order of magnitude, or even ranked oppositely (the intra-state contribution is bigger than the inter-state).

These results show that the LR representations obtained with single linkage clustering do not fully capture the information contained in the data and destroy a comparable, or even larger, amount of information than what is maintained. In general, the inter-state (resp. intra-state) contribution for average linkage is always significantly higher (resp. lower) than that obtained with single linkage, and the outcomes of the two linkage methods are comparable only when more than half of the frames are retained. Results from dimensionality reduction also support the observation that the average linkage identifies more informative LR representations than those produced by single linkage in that, coherently with the trend of MSR, the PCC value is consistently higher in the former case than in the latter.

In the three panels of \hyperref[fig:1ake-PCC-cov]{Fig.\ref{fig:1ake-PCC-cov}} the yellow stars indicate the representations that, for each method, maximise the relevance. Interestingly, in both the graph of covariance matrices and of the PCC, these representations are at the elbow of the curve. This behaviour is suggestive of the fact that further increases in resolution lead to an increased model complexity that is no longer balanced by information gain: the tradeoff between complexity and informativeness turns in favour of the former, consistently with the interpretation of relevance as a measure of useful information content.

Finally, we compare the data obtained for average and single linkage with those of the random clustering. The latter has a very close relevance curve to that of single linkage, and the MSR values associated with single linkage ($MSR_S=0.235$) and random clustering ($MSR_R=0.233$) differ only at the third decimal place; in spite of that, the informativeness of the partitions obtained with an information-driven protocol is incomparably greater than that returned by random clustering: the trace of the inter-cluster covariance matrix of the latter is always lower than the intra-cluster one until we consider representations in which about $2/3$ of the original frames are preserved, and the PCC between reference and random partition DC is almost zero at any level of resolution. These observations further support the idea that the relevance alone cannot be taken as an absolute measure of the informativeness of a given low-resolution representation, however this quantity in combination with the appropriate classification method proves extremely effective in identifying protocols that maximises the emergence of useful information.

Finally, we looked in detail at the three representations that maximise the relevance for each of the clustering methods under examination. In the right-hand side of \hyperref[fig:1akeDC]{Fig.\ref{fig:1akeDC}} the distributions of centroids in LR representations are compared with the frame distributions in HR ones, as the points in each panel are coloured according to the value taken from the first DC in the LR representation. A visual inspection of these data shows that the distribution of average linkage centroids in LR DC space is consistent with that of the HR frames; in both graphs it is possible to recognise a colour gradient along the $x$ axis, showing that neighbouring frames in the HR diffusion space are grouped together in the LR space. As for the linkage criterion, the LR representation maximising the relevance produces a slightly different distribution of points than that of the HR frames; furthermore, looking at the colour of points in both spaces it appears that distant frames in the diffusion space are associated to the same cluster. This is even more evident when correlating the values assumed by the DCs in the HR and LR representation, as shown in the bottom-left corner of \hyperref[fig:1akeDC]{Fig.\ref{fig:1akeDC}}. For both linkage measures (average and single) it is possible to identify a strong correlation between the first DCs in LR and HR: the Pearson correlation coefficient is $0.95$ for average linkage and  $0.85$ for single linkage; nevertheless, in the case of single linkage, some clusters contain frames with a wide distribution of HR diffusion coordinate values, i.e., frames carrying very different information are mistakingly lumped in the same bin. Last but not least, we observe, as expected, a total lack of correlation - both in terms of point distribution and cluster composition - between the DCs of the random partition and the reference HR DCs.

\section{Conclusions}

The steady increase of available computational power offers impressive opportunities in the investigation of biological macromolecules; at the same time, however, the corresponding growth of the pile of data produce by {\it in silico} studies requires the application of coarse-graining and dimensionality reduction techniques that allow one to discriminate between signal and noise in the dataset, and extract simple, useful, and intelligible information out of it. To this end, the resolution-relevance framework represents a novel, powerful instrument to construct informative simplified representations of a molecule's conformational space; however, a blind and black-box application of this approach bears the risk of giving high-relevance partitions more credit than they deserve, in that the quality of said partitions cannot be disentangled by the specific classification method employed to construct them. In the present work we have tackled this issue through the systematic, dataset-wide application of the resolution-relevance framework to a number of structurally distinct proteins, making use of many of the most diffused and commonly used clustering methods. Our results show that the clustering strategies, and more specifically the particular definitions of inter-cluster distance, employed to group together ``similar'' frames into structurally homogeneous clusters return different values of the multi-scale relevance (MSR), a global measure of the relevance at various levels of resolution. We find that the partitions having higher values of the MSR are those that produce the most physically sensible partitions, as quantified in terms of intra- and inter-cluster covariance, as well as the correlation between the collective diffusion coordinates computed in the reference, high-resolution space and those of the low-resolution representation. Most interestingly, a positive correlation emerges between high values of MSR and the efficacy of a clustering method in reconstructing a low-resolution representation that features an ultrametric structure: this observation is suggestive of the fact that the configurational space spanned by a protein in the course of a molecular dynamics simulation is intrinsically organised in a hierarchical manner, which is consistent with the hypothesis, proposed and verified in the literature, that the free energy landscape of proteins is effectively self-similar.

In conclusion, we propose that the clustering method employed in the dimensionality reduction of a dataset could be not only employed as a tool to preprocess the data {\it in order to analyse them}, but also treated as an analysis tool itself: in fact, through the joint usage with the general, parameter-free resolution-relevance framework it is possible to discriminate among partitioning approaches that produce low-resolution models more or less representative of the salient qualities of the high-resolution reference. The combination of these algorithms can thus pave the way to an even more fruitful deployment of clustering approaches in computational biophysics, bringing further insight in the behaviour of complex macromolecules.
\section{Methods}
\label{sec:Methods}
\subsection{Protein Selection}
\label{subsec:PS}

For the exploratory analysis it was essential to employ a set of structurally distinct and uncorrelated proteins, in order to draw general conclusions. To this end, a dataset of $107$ proteins, including many of the known folds and structure classes, was constructed and clustered based on their dynamics. For each protein, the first $10$ normal modes of fluctuation were analysed using an elastic network model~\cite{micheletti2004accurate}, and superimposed by means of the ALADYN~\cite{potestio2010aladyn} protocol, which performs a hybrid structural/dynamical alignment. The similarity between the essential spaces spanned by the first $10$ normal modes was quantified by means of the root mean square inner product~\cite{amadei1999convergence} (RMSIP). The distance between the essential dynamics of two aligned proteins was defined as $d_{ij} = 1-$ RMSIP$_{ij}$; this distance was employed to perform a hierarchical clustering. The resulting dendrogram allowed us to identify $12$ clusters, each of which contains proteins whose dynamics are similar (RMSIP above $0.5$). The $12$ proteins used in this work are the centroids of these clusters, and their PDB codes are: 1DSL, 1NOA, 1SNO, 1UNE, 1XWL, 1IGD, 1HYP, 2FGF, 1KNT, 1QKE, 2EXO, 1KOE.

Two specific proteins were used for the second part of the analysis. The protein adenylate kinase (PDB code AKE4) because of its relatively small size and the possibility to observe conformational transitions over time scales easily achievable by means of plain MD. The second system is the humanised IgG4 monoclonal antibody (PDB code 5DK3). This system was chosen because of its large size and higher structural and dynamical complexity~\cite{tarenzi2021communication}. As the results obtained in the two cases are consistent, for the sake of clarity we only reported the data pertaining the adenylate kinase in the main text, while those of the antibody are provided as Supporting information.

\subsection{Simulation setup}
\label{subsec:Setup}

For all biological systems studied, the Gromacs 2018~\cite{bekker1993gromacs,abraham2015gromacs} software was employed, and the topology was defined through the AMBER99SB-ILDN~\cite{lindorff2010improved} force field. The simulations were performed in explicit solvent, the latter being TIP3P water~\cite{hockney2021computer}; $Na^+$ and $Cl^-$ ions at the concentration of  $0.15\,M$ were added to neutralise the global electric charge and mimic physiological ion concentration in the cell. Energy minimisation was performed until the maximum force reached a specific value, $F_{min} = 1000\ kJ mol^{-1}nm^{-1}$ for the 12-protein dataset and $F_{min} = 500\ kJ mol^{-1}nm^{-1}$ in case of the adenylate kinase protein and the humanised IgG4 monoclonal antibody. NVT and NPT equilibrations were performed using the velocity-rescale thermostat~\cite{bussi2007canonical} and the Parrinello-Rahman barostat~\cite{parrinello1981polymorphic}. With respect to the interaction, a cut-off was used for Van der Waals interaction and for the short-range component of the Coulomb one. The long-range component of the Coulomb force, instead, was computed with the Particle Mesh Ewald algorithm. The LINCS algorithm~\cite{hess1997lincs} was employed to define the constraints on the hydrogen-containing bonds and allows an integration time of $2\,fs$. The dynamic of each system in the 12-protein dataset has been simulated for $300\;ns$, regarding the adenylate kinase and the humanised IgG4 monoclonal antibody their dynamic were simulated for $800\;ns$ and $2\;\mu s$, respectively.

\subsection{Clustering}
For each biological system the RMSD matrix was computed~\cite{mcgibbon2015mdtraj} for three atom selection: all atoms, C$_\alpha$ only, C$_\beta$ only. From each of these matrices, $7$ dendrograms were constructed exploring the $7$ different definitions of inter-cluster distance supported by the python module Syipy~\cite{leman2020macromolecular}, employed for the clustering analysis. Each dendrogram was cut at different levels, ranging the number of clusters from $1$ to the number of frames $M$ by steps of $10$. For each resulting partition, the corresponding resolution and relevance values were computed following \hyperref[eq:rel]{Eq.\ref{eq:rel}} and \hyperref[eq:res]{Eq.\ref{eq:res}}. In this way, for each system $21$ different curves are obtained: $3$ representations of the system (all, C$_\alpha$, C$_\beta$) times $7$ measures of distance between clusters.

\subsection{Multi-scale relevance and random Curves}
\label{subsec:random}

In addition to information-driven clustering, random clustering is also possible. Given a number of clusters $K$, a label vector can be randomly generated by iteratively sampling $M$ elements from a list containing all integers from $1$ to $K$. For a given $K$, $10^4$ vectors of labels were generated, and for each of them the corresponding values of relevance and resolution were calculated. The points on the random curve were obtained by averaging the relevance and resolution values obtained for a given number of clusters $K$, and varying the number of clusters from $1$ to $M$ by steps of $10$. The MSR value associated to this curve was used to normalise the MSRs resulting from the hierarchical clustering algorithms:
\begin{equation}
    \overline{MSR}_{i}=\frac{MSR_{i}-MSR_R}{MSR_R}
    \label{eq:MSR-norm}
\end{equation}
where $MSR_i$ is the area under the relevance-resolution curve drawn by algorithm type $i$, and $MSR_R$ is the one obtained by the random procedure.

\subsection{Diffusion Maps}
\label{subsec:DM}

Diffusion Maps is a nonlinear dimensionality reduction tool that rearranges the dataset according to its connectivity~\cite{DiffMap1,nadler2006diffusion,lafon2006diffusion}. The connectivity of the data points is  measured using a local similarity measure, and is used to create a time-dependent diffusion process. The algorithm was employed to compare the manifold obtained from the whole trajectory (\textit{high-resolution} representation) and those obtained considering only the centroids of some partitions (\textit{low-resolution} representation). In the high-resolution representation, the inter-frame distance is the RMSD matrix, as for clustering. In the low-resolution one, each cluster is represented by its centroid and the distance between clusters is given by the linkage measure adopted. The terms \textit{high-resolution} and \textit{low-resolution} are used here in connection with the number of frames retained and not to the selection of atoms, which is the C$_\beta$ atom selection in both scenarios. The algorithm also requires a threshold parameter to determine what is \textit{near} or \textit{far} in the source data set. To be consistent in its approach, the quantile of order $0.1$ of the distribution of distances was always chosen. It is possible to relate the points of the HR space to those of the LR one, since on one side there are the elements of the clusters, and on the other the centroids. To make this visual inspection easier, data were rescaled so that all points were contained in the square $[-1,1]^2$.

\subsection{Covariance Matrix}
\label{subsec:CM}

The covariance matrix of the positions of the C$_\beta$ atoms along the trajectory can be subdivided in two contributions: inter- and intra-clusters~\cite{pontiggia2008small,kitao1998energy, pontiggia2007anharmonicity}. The correlation between two elements of the system (in this case two atoms C$_\beta$) is given by:
\begin{align}
    &C_{ij}=C_{ij}^{intra}+C_{ij}^{inter} \label{eq:cov1}\\
    &C_{ij}^{intra}=\sum_l \omega_l  \langle [ \Vec{r}_i-\langle \Vec{r}_i\rangle_l][ \Vec{r}_j-\langle \Vec{r}_j\rangle_l ]\rangle_l \label{eq:cov2}\\
    & C_{ij}^{inter}=  \sum_l \omega_l [\langle\Vec{r}_i \rangle_l - \langle\Vec{r}_i\rangle ]  [\langle\Vec{r}_j \rangle_l - \langle\Vec{r}_j\rangle ] \label{eq:cov3}
\end{align}
where $l$ runs over a cluster $C_l$; $\omega_l$ is the weight of the state $l$, which is the fraction of simulation time spent by the system in it; $\langle \rangle_l$ denotes the average over the conformations belonging to the cluster $l$. The decomposition is performed in analogy with the ``jumping among minima'' model~\cite{kitao1998energy}. The first term in the decomposition of the covariance matrix is the contribution arising from structural fluctuations within clusters. The second term arises from the structural differences of the clusters centroids. Consequently, a good partition will have a high inter-cluster term and a low intra-cluster one.


\section*{Appendix}
\label{app:linkage}

\begin{itemize}
    \item \textbf{Single linkage}: the distance between a pair of clusters is determined by the two closest objects belonging to the different clusters.
    \begin{eqnarray}
        D(C_i,C_j)&=&\min\{d(\textbf{x}_i,\textbf{x}_j)\\ \nonumber
        &\forall& \; \textbf{x}_i \in C_i\; and \; \textbf{x}_j \in C_j  \}
    \end{eqnarray}
    Single linkage clustering tends to produce elongated clusters, which causes the \textit{chaining effect}. Two points that form a bridge between two clusters cause the single-link clustering to join these two clusters into one.
    \item \textbf{Complete linkage}: it consider the distance between two clusters to be equal to the largest distance from any member of one cluster to any member of the other cluster.
    \begin{eqnarray}
        D(C_i,C_j)&=&\max\{d(\textbf{x}_i,\textbf{x}_j)\\ \nonumber
        &\forall& \; \textbf{x}_i \in C_i\; and\; \textbf{x}_j \in C_j  \}
    \end{eqnarray}
    This procedure tends to form smaller and more compact clusters.
    \item \textbf{Average linkage}: it considers the distance between two clusters as the average distance between all pairs of points coming from the different groups.
    \begin{equation}
         D(C_i,C_j)=\frac{1}{|C_i|\,|C_j|}\sum_{\textbf{x}_i \in C_i}\; \sum_{\textbf{x}_j \in C_j}  d(\textbf{x}_i,\textbf{x}_j)
    \end{equation}
    where $|.|$ stands for the cardinality of set, i.e. the number of items pertaining to it. This approach can cause the splitting of elongated clusters and the merging of portions of neighbouring elongated clusters.
    \item \textbf{Weighted linkage}: also in this case, the protocol takes as cluster distance the average distance from any member of one cluster to any member of the other one. The difference is that the distance between the new cluster and another is weighted with respect to the number of data in each cluster. Consequently, the distance between the cluster $C_k=C_i\cup C_j$ and a third cluster $C_l$, not involved in the definition of $C_k$, is:
    \begin{align}
        D(C_k,C_l)&=\frac{1}{2|C_i||C_l|}\sum_{\textbf{x}_i \in C_i}\sum_{\textbf{x}_l \in C_l}d(\textbf{x}_i,\textbf{x}_l)+\nonumber \\[1em]
        &+\frac{1}{2|C_j||C_l|}\sum_{\textbf{x}_j \in C_j}\sum_{\textbf{x}_l \in C_l}d(\textbf{x}_j,\textbf{x}_l)
    \end{align}
    \item \textbf{Centroid linkage}: in this case, two clusters are merged based on the distance of their centroids. The definition of centroids is:
    \begin{equation}
        \mu_i= \frac{1}{|C_i|}\sum_{\textbf{x}_i\in C_i} \textbf{x}_i
        \label{eq:centroid1}
    \end{equation}
    Consequently, the distance between clusters results the Euclidean distance between the centroids:
    \begin{equation}
        D(C_i,C_j)= ||\mu_i-\mu_j||_2
    \end{equation}
    The centroid of the resulting cluster $C_k=C_i\cup C_j$ is recomputed according to \hyperref[eq:centroid1]{Eq.\ref{eq:centroid1}} considering all the points belonging to it.
    \item \textbf{Median linkage}: the procedure is similar to the centroid linkage, except that the centroid of the resulting cluster $\mu_k$ is the average of the centroid of the merged ones:
    \begin{equation}
        \mu_k=\frac{1}{2}(\mu_i+\mu_k)
    \end{equation}
    This is equivalent to giving the same weight to merged clusters regardless of the number of elements in them.
    \item \textbf{Ward linkage}: the methods aims to minimise the increase of the intra-cluster sum of squared errors:
    \begin{equation}
        E=\sum_{k=1}^K\sum_{\textbf{x}_i\in C_k} ||\textbf{x}_i-\mu_k||^2_2
    \end{equation}
    where $K$ is the number of clusters and $\mu_k$ is the centroid of the cluster $C_k$ (defined by  \hyperref[eq:centroid1]{Eq.\ref{eq:centroid1}}). Merging clusters $C_i$ and $C_j$ produces an increase in variance of
    \begin{equation}
        \Delta E= \frac{n_in_j}{n_i+n_j}||\mu_i-\mu_j||^2_2.
    \end{equation}
    Consequently, the distance between the new cluster $C_k=C_i \cup C_j$ and an unused cluster $C_l$ is given by the recursive equation:
    \begin{align}
        D(C_l,C_k)=&\frac{|C_i|+|C_l|}{|C_i|+|C_j|+|C_l|}D(C_l,C_i)+\nonumber\\[1em]
        &+\frac{|C_j|+|C_l|}{|C_i|+|C_j|+|C_l|}D(C_l,C_j)+\nonumber\\[1em]
        &-\frac{|C_l|}{(|C_i|+|C_j|)^2}D(C_i,C_j)
    \end{align}
\end{itemize}

All definitions of distance between clusters can be summarised by the recursive relation proposed by Lance and Williams~\cite{lance1967general}:
\begin{align}
    D(C_l, C_i\cup C_j)=& \alpha_i D(C_l,C_i)+ \alpha_j D(C_l,C_j)+\beta D(C_i,C_i) + \nonumber \\[1em]
    &+\gamma |D(C_l,C_i)-D(C_l,C_j)|
    \label{eq:recursive}
\end{align}
where $\alpha_i,\;\alpha_j,\;\beta$ and $\gamma$ are coefficients that takes different values depending on the protocol used.

\section*{Author Contributions}
RP conceptualised the study. RP and RC supervised the study. MM developed the methodology, performed the simulations, wrote the analysis software, and wrote the original draft. MM, RP and RC wrote, reviewed, and edited the final manuscript.

\section*{Conflicts of interest}
There are no conflicts to declare.

\section*{Data availability}
The raw data produced and analysed in this work are freely available on the Zenodo repository with DOI \url{10.5281/zenodo.6554498}.

\section*{Acknowledgements}
The authors are indebted with Thomas Tarenzi for valuable help in the construction of the protein dataset, and with Luca Tubiana for a critical and insightful reading of the manuscript. This project received funding from the European Research Council (ERC) under the European Union's Horizon 2020 research and innovation program (Grant 758588).




\renewcommand\refname{\large References}

\bibliography{main.bbl}

\end{document}